\documentclass[12pt]{article}
\newcommand{\be}{\begin{equation}}
\newcommand{\bea}{\begin{eqnarray}}
\newcommand{\eea}{\end{eqnarray}}
\newcommand{\ba}{\begin{array}}
\newcommand{\ea}{\end{array}}
\newcommand{\ee}{\end{equation}}

\expandafter\ifx\csname mathbbm\endcsname\relax

\else

\fi \textheight 22cm \textwidth 15cm \topmargin 1mm
\begin{document}
\begin{titlepage}
\hfill \vbox{
    \halign{#\hfil         \cr
           IPM/P-2005/052 \cr
           hep-th/0509032  \cr
           } 
      }  
\vspace*{20mm}
\begin{center}
{\Large {\bf DBI with Primordial Magnetic Field in the Sky }\\
}

\vspace*{15mm} \vspace*{1mm} {Mohammad A. Ganjali}
 \\
\vspace*{1cm}

{\it Institute for Studies in Theoretical Physics
and Mathematics (IPM)\\
 \vspace{3mm}
 Department of Physics, Sharif University of Technology\\
P.O. Box 11365-9161, Tehran, Iran}\\
\vspace*{.5cm} E-mail:Ganjali@theory.ipm.ac.ir\\

 \vspace*{1cm}
\end{center}
\begin{abstract}
In this paper, we study the generation of a large scale magnetic
field with amplitude of order $\mu$G in an inflationary model
which has been introduced in hep/th 0310221. This inflationary
model based on existence of a speed limit for inflaton field.
Generating a mass for inflaton at scale above the $\phi_{IR}$,
breaks the conformal triviality of the Maxwell equation and causes
to originate a magnetic field during the inflation. The amplitude
strongly depends on the details of reheating stage and also
depends on the e-foldings parameter N. We find the amplitude of
the primordial magnetic field at decoupling time in this
inflationary background using late time behavior of the theory.
\end{abstract}

\end{titlepage}
\section{Introduction}
The observational data of galaxies clusters, disks and spiral
galaxies displays a large scale magnetic field with coherence
scale of few kpc and an amplitude of order $\mu$G
$(\cite{Bassett:2000aw}-\cite{Davis:2005ih})$. Such amplitudes
present an "inverse" fine tuning problem as compared with the
standard one in inflation \cite{Bassett:2000aw}.

Although, the origin of such seed field is not exactly known, but
several mechanism for originating the magnetic field have been
supposed and some limits on its amplitude has  been found.
Magnetic field can be produced by charge separation mechanism
during galaxy formation, or by amplification of preexisting seed
field. In the second case, the amplification can be achieved
either by the adiabatic compression of the fields in the collapse
of the protogalactic cloud, or by the galactic dynamo mechanism
where the differential rotation of the galaxy is able to transfer
kinetic energy into magnetic field. In the latter case, the limit
on the primordial seed fields at the decoupling time is in the
range $B_{dec}\sim10^{-17}-10^{-20}G$ \cite{Davis:1999bt} for flat
universe without cosmological constant. For a flat universe with
cosmological constant the limits are relaxed up to
$B_{dec}\sim10^{-25}-10^{-30}G$ \cite{Maroto:2001ki}. Such seed
field is amplified exponentially by galactic dynamo, or other
mechanism, until it reaches to the observed value $B_{dec}\sim
10^{-6}G$.

However, any homogenous magnetic field existing before decoupling
time should be weaker than $B_0\leq10^{-9}G$ in order to avoid the
production of large anisotropies in the cosmic microwave
background radiation \cite{Maroto:2001ki}.

All the above mechanism need a primordial seed field. There are
two main explanation for generating such a field. On the one hand,
those based on the phase transition in the early universe. The
smallness of the field coherence is the main problem of these
theories. On the other hand, the electromagnetic quantum
fluctuation during inflation is another source for amplifying the
field\cite{Turner:1987bw}. The main problem in this case is that
in order to get some amplification it is necessary to breaks the
conformal symmetry of the Maxwell equation in
Friedmann-Robertson-Walker background. In fact, there is no mixing
between positive and negative frequency modes of the solution of
Maxwell equation, hence there is no photon production
\cite{Maroto:2001ki}. Therefore, in order to get magnetic field
amplification we need to break conformal invariance.

There are several possible sources for breaking the conformal
invariance for example: non zero masses of charged particles,
absence of conformal invariance for even massless scaler field,
quantum conformal anomaly due to the triangle diagrams, and
breaking the $U(1)$ gauge invariance as well as conformal
invariance\cite{Dolgov:1993vg}.

Although, the generation of magnetic field in the standard model
of cosmology was studied in various frameworks it was also studied
in string cosmology
$(\cite{Gasperini:1995dh},\cite{Garousi:2004hy},
\cite{Bamba:2004cu},\cite{Davis:2005ih})$. String cosmology
provides a very natural way for breaking the conformal triviality
of the Maxwell equation. In fact, the story was started when the
author of \cite{Giddings:2001yu} found a warped geometry in the
presence of fluxes and branes. By adding anti$-D3$ brane to these
solution, one can lift the $ADS$ vacuum to $DS$ vacuum and find
locally stable minimum $($KKLT vacua$)$. So, it is very natural to
study inflation in string theory by considering the KKLT vacua.
Many different efforts was done to finding an inflationary model
in string theory in which is in agreement with observational data
$(\cite{Silverstein:2003hf},\cite{Alishahiha:2004eh}
,\cite{Kachru:2003aw},\cite{Kachru:2003sx})$.

Furthermore, one may suppose the fluctuation of gauge field via
DBI action in a typical  inflationary background and study the
dynamic of such field. In inflationary model which we are
interested
in$(\cite{Silverstein:2003hf},\cite{Alishahiha:2004eh})$,
conformal invariance breaks in some region and magnetic field can
be produced in the early universe.

In the following subsection, we will present a preview of this
inflationary model which based on D-cceleration mechanism for
inflation, then in section $2$ we'll study the equation of motion
of the gauge field and study it by using the late time behavior of
various parameters and fields in this inflationary background and
find the amplitude for the magnetic field in term of e-foldings
parameter. Some comments about reheating stage will be presented
in section 3.

\subsection{Review of D-cceleration}
In this model, "D-cceleration" mechanism in strongly coupled
conformal field theories causes for slowing motion of scalar field
$(\cite{Silverstein:2003hf},\cite{Alishahiha:2004eh})$. The
dynamic of the field $\phi$ in strongly 't Hooft coupled conformal
field theory is easy understood in gravity side of the AdS/CFT
correspondence\cite{Maldacena:1997re}.

In CFT side, the slowdown arising from the virtual effects of the
light particles. Moreover, the integrating out of such $\chi$
fields produces higher derivative terms in the effective action
which are captured in DBI action and due to speed limit on rolling
down of scalar field.

In gravity side, a probe $D_3$ brane travelling down a
five-dimensional warped throat and  the inflaton field $\phi$ is a
scalar parameterizing a direction on the approximate Colomb branch
of the system.  Motion toward the origin is motion in the radial
direction of the $AdS_5$, toward the horizon where new degrees of
freedom $\chi$ becomes light. The motion of $\phi$ is constrained
by the causal speed limit on the gravity side, leading to slow
roll even in the presence of a steep potential.

The dynamic of the probe $D_3$ brane coupled to gravity is given
by DBI action $(\cite{Aharony:1999ti},\cite{Silverstein:2003hf})$,
      \bea \label{f1}
           S=\int \frac{1}{2}M_P^2\sqrt{-g}\Re +\pounds_{eff}+...,
      \eea
where
      \bea  \label{f2}
         \pounds_{eff}=-\frac{1}{g_s}\sqrt{-g}\big{(}f(\phi)^{-1}
         \sqrt{1+f(\phi)g^{\mu\nu}\partial_\mu\phi\partial_\nu\phi+
         \sqrt{f}g^{\mu\nu}F_{\mu\nu}}+V(\phi)\big{)}.
      \eea
Here $F_{\mu\nu}=\partial _{[\mu}A_{\nu]}$. The function $f(\phi)$
is the $(squared)$ warped factor of the AdS-like throat. For
example, for a pure $AdS_5$ of radius R, it is
$f(\phi)=\frac{\lambda}{\phi^4}$ with
$\lambda\equiv\frac{R^4}{\alpha'^2}$.

The potential $V(\phi)$ arises from the coupling of the $D$ brane
to background RR flux. In the case of ${\cal N}=4$ theory, it is
quartic and conformal invariance prohibits the generation of a
mass term for $\phi$. But, for this approximate AdS throat,
conformal invariance holds only at scale above $\phi_{IR}$ and
there is nothing preventing the generation of a mass for $\phi$ in
which
    \bea  \label{f3}
                V(\phi)=m^2\phi^2.
    \eea
This leads to a $\chi$ mass of order $\sqrt{g_s}m$ cutting off the
throat, so the mass satisfies $m\leq\frac{\phi_{IR}}{\sqrt{g_s}}$.
One can also add cosmological constant term to the action $(\cite
{f1})$ using KKLT vacua.

 Considering the flat FRW metric,
    \bea  \label{f4}
         ds^2=-dt^2+a(t)^2\delta_{ij}dx^idx^j
    \eea
and using the Hamilton-Jacobi approach in which the field $\phi$
roles as cosmological time, one obtains
$(\cite{Silverstein:2003hf},\cite{Alishahiha:2004eh})$,
     \bea   \label{f5}
         \dot{\phi}&=&-2g_sM_P^2\frac{H'(\phi)}{\gamma(\phi)},\cr
         V(\phi)&=&3g_sM_P^2H(\phi)^2-\gamma(\phi)/f(\phi)+1/f(\phi),
     \eea
where
     \bea   \label{g5}
        \gamma(\phi)&=&\frac{1}{\sqrt{1-f(\phi)\dot{\phi}^2}}
         =(1+4g_s^2M_P^2f(\phi)H'(\phi)^2)^{1/2},
     \eea
and the slow roll parameter is
    \bea   \label{f6}
            \epsilon=\frac{2g_sM_P^2}{\gamma}(\frac{H'}{H})^2.
    \eea
The late time behavior of the scale factor $a(t)$ is
    \bea   \label{f7}
            a(t)\rightarrow a_0 t^{1/\epsilon}
    \eea
We see that for $\epsilon<1$ we have a power low inflation. A
detailed analysis shows that
    \bea   \label{f8}
              \frac{1}{\epsilon}=
              \frac{1}{3}(1+\sqrt{1+\frac{3m^2\lambda}{g_sM_P^2}})
              \simeq\sqrt{\frac{\lambda}{3g_s}}\frac{m}{M_P}
    \eea
and the condition $\epsilon<1$ means that $m^2\lambda>g_sM_P^2$.
In the other hand, accelerated expansion occurs only if the mass
of inflaton $m$ is suitably large compared with Planck mass.

The number of e-foldings parameter is given by
     \bea    \label{f9}
           N=\frac{1}{\epsilon}\log{(\frac{\phi_i}{\phi_e})},
     \eea
where $\phi_i$ and $\phi_e$ are the inflaton field at the begin
and the end of the inflation respectively. There are possibilities
to start and end inflation anywhere within the ADS-like geometry,
subject to $\phi_{IR}\leq\phi_{end}<\phi_{start}\leq\phi_{UV}$,
where $\phi_{IR}$ and $\phi_{UV}$ are determined by the geometry
of the throat. The UV end of the throat arises at the scale
$\phi_{UV}\alpha'\sim R$ or
$\frac{\phi_{UV}}{\sqrt{g_s}}\sim\frac{\lambda^{1/8}M_P\sqrt{g_s}}{\sqrt{V_X}}$,
where $V_X$ is the volume in string unit of $X$. Note also that
$\phi_{IR}>\sqrt{g_s}m$.
\section{DBI with Magnetic Fluctuation}
The D-3 brane dynamic in the ADS geometry is described at low
energy by DBI action $(\ref{f2})$, which is the Yang-Mills action
plus some higher derivative corrections. Considering $(\ref{f2})$
and using the expansion
    \bea  \label{f10}
      \sqrt{det(M_0+M)}&=&\sqrt{det(M_0)}[1+\frac{1}{2}Tr(M_0^{-1}M)-
      \frac{1}{4}Tr(M_0^{-1}M)^2\cr &+&\frac{1}{8}(Tr(M_0^{-1}M))^2
      +\frac{1}{6}Tr(M_0^{-1}M)^3\cr &-& \frac{1}{8}Tr(M_0^{-1}M)Tr(M_0^{-1}M)^2
      +\frac{1}{48}(Tr(M_0^{-1}M))^3+...],
     \eea
one finds the final action up to second order of gauge field as
       \bea   \label{f11}
           -\frac{1}{g_{YM}^2}\int d^4x[\gamma^{-1}(\frac{1}{2a}
           (\bigtriangledown\times\overrightarrow{A}).
           (\bigtriangledown\times\overrightarrow{A})
           -\frac{a}{2}\dot{\overrightarrow{A}}^2\gamma^2)
           +f^{-1}(\gamma^{-1}-1)+V(\phi)].
       \eea
Here we used the Coulomb gauge $(A_0=0=\vec{\nabla}.\vec{A})$. The
variation of the action over the field $\vec{A}$ gives us the
following equation of motion
      \bea   \label{f12}
         \ddot{\overrightarrow{A}}+\dot{\overrightarrow{A}}
         [H+\frac{\gamma^2}{2}
         (f'\dot{\phi}^3+2f\dot{\phi}\ddot{\phi})]
         -\frac{1}{\gamma^2a^2}\nabla^2\overrightarrow{A}=0,
      \eea
Using the standard trick, which is briefly explained at appendix,
this differential equation can be written as
     \bea  \label {g12}
         \tilde{\overrightarrow{A}}''+[k^2-f(\eta)]\tilde{\overrightarrow{A}}=0,
     \eea
where $\eta=\int a^{-1}dt$ is the conformal time and new variable
$\tilde{\overrightarrow{A_k}}$ is defined as
$\overrightarrow{A}_k\equiv
b^{\frac{-1}{2}}\tilde{\overrightarrow{A_k}}$ and we have
     \bea    \label{f13}
          f(\eta)&=&k^2f\dot{\phi}^2+\frac{a^2}{4}[(F^2-H^2)+2(\dot{F}-\dot{H})]\cr
          F&=&H+\frac{\gamma^2}{2}
         (f'\dot{\phi}^3+2f\dot{\phi}\ddot{\phi}),
     \eea
The solution for equation $(\ref{g12})$ is
     \bea   \label{f14}
         \tilde{\vec{A}_k}=\alpha_k\tilde{\vec{A}_k^i}+
         \beta_k\tilde{\vec{A}_k^{\ast i}},
     \eea
where the Bogolyubov coefficient is
     \bea   \label{f15}
         \beta_k=-i\int_{\eta_i}^{\eta_f}\tilde{\vec{A}_k^i}
         f(\eta')\tilde{\vec{A}_k}d\eta'.
     \eea
The quantities $ \tilde{\vec{A}}^i_k $ are the initial value for
the field $\tilde{\vec{A}_k}$. It is easy to see that at
asymptotic past time $\eta \rightarrow -\infty$ the $f(\eta)$ term
goes to zero and the solution is
$\tilde{\vec{A}_k^i}=e^{-ik\eta}/\sqrt{2k}$.

The energy density stored in a magnetic field is given by
     \bea   \label{f16}
          \rho_B=\frac{|B_k^2|}{8\pi}.
     \eea
We want to find the amplitude of magnetic field at scale
corresponds to decoupling time,
$\omega_{dec}=\frac{k_{dec}}{a_{dec}}$. Then
      \bea   \label{f17}
        |B_k|=\sqrt{8\pi}(\omega_{dec})^2|\beta_k|.
      \eea
In this step, one may use the procedure that one applied  in
\cite{Garousi:2004hy}, by defining a dimensionless quantity
$\alpha$. Similarly, but with small changes, we want to find the
amplitude by finding the late time behavior of the $f(\eta)$.

\subsection{Evaluation of $f(\eta)$}
Let us compute the late time behavior of $f(\eta)$ using
$(\ref{f13})$. The equation of motion for $\phi$, by neglecting
the gauge field effects, is \cite{Silverstein:2003hf}
     \bea   \label{f18}
         \ddot{\phi}+\frac{3f'}{2f}\dot{\phi}^2-\frac{f'}{f^2}+\frac{3H}{\gamma^2}\dot{\phi}
         +(V'+\frac{f'}{f^2})\frac{1}{\gamma^3}=0.
     \eea
The late time behavior of scale factor is power law as
$a=a_0t^{1/{\epsilon}}$. In this case, considering slow roll
condition $\gamma\gg 1$, $($which means that$ f \sim
\dot{\phi}^2$$)$  and using
$(\ref{f5},\ref{f7},\ref{f13},\ref{f18})$ and the late time
behavior of $\phi,\gamma, H$ as
      \bea   \label{f19}
         \phi\rightarrow\frac{\sqrt{\lambda}}{t},\;\;\;\;\;\;
         \gamma\rightarrow
         \sqrt{\frac{4g_s}{3\lambda}}M_Pmt^2,\;\;\;\;\;\;
         H\rightarrow\frac{1}{\epsilon t},
      \eea
one finds following expression for the late time behavior of $f$
in term of conformal time $\eta$,
      \bea   \label{f20}
         f(\eta)\rightarrow k^2+\Sigma_{\beta=1}^{\beta=5}f_{\beta}h_{\beta}
         \eta^{\frac{-2(1-\beta\epsilon)}{1-\epsilon}},
      \eea
where
      \bea   \label{f22}
            h_{\beta}=(ma_0^{\frac{\epsilon}{\epsilon-1}})^{-2(\beta-1)}.
      \eea
and $f_{\beta}$'s are functions of $\epsilon$ as
      \bea    \label{f21}
        f_1&=&(\frac{\epsilon-1}{\epsilon})^{\frac{2(1-\epsilon)}{\epsilon-1}}
        [-\frac{24\epsilon+3}{2\epsilon^2}
        -\frac{7\epsilon+1}{2\epsilon^2}(9-6\epsilon)^{1/2}],\cr
        f_2&=&(\frac{\epsilon-1}{\epsilon})^{\frac{2(1-2\epsilon)}{\epsilon-1}}
        [\frac{9\epsilon+1}{2\epsilon^2}(9-6\epsilon)^{1/2}
        +\frac{9-6\epsilon}{\epsilon^2}]\cr
        f_3&=&(\frac{\epsilon-1}{\epsilon})^{\frac{2(1-3\epsilon)}{\epsilon-1}}
        [\frac{-8\epsilon^2-6\epsilon+27}{8\epsilon^2}(9-6\epsilon)+\frac{9}{8\epsilon^4}
        (9-6\epsilon)^{3/2}],\cr
        f_4&=&(\frac{\epsilon-1}{\epsilon})^{\frac{2(1-4\epsilon)}{\epsilon-1}}
        [-\frac{9}{4\epsilon^4}(9-6\epsilon)^{3/2}
        -\frac{(9-6\epsilon)^2}{2\epsilon^4}],\cr
        f_5&=&(\frac{\epsilon-1}{\epsilon})^{\frac{2(1-5\epsilon)}{\epsilon-1}}
        [\frac{(9-6\epsilon)^2}{2\epsilon^4}].
      \eea
The constant $a_0$ which arises in time evaluation of scale factor
is a model dependent quantity. In fact, as it was mentioned in
\cite{Alishahiha:2004eh}, the initial value for $\phi$ at starting
point is constrained within the consistence between predicted
non-gaussianity of the theory and non-gaussianity data from WMAP
analysis. For example, for the case $\epsilon=1/20$, which is our
interested case, the initial value for inflaton field is
$\phi_i\sim 1.5 \sqrt{g_s}M_P$\cite{Alishahiha:2004eh}. From this,
the amount of the $a_0$ depends on details of the reheating, the
value of $\phi$ at starting time and at the end of the inflation.
In the other word, the e-folding parameter has a crucial roll in
the $a_0$. Considering $(\ref{f9},\ref{f22})$, one obtains
following $a_0$
    \bea   \label{f23}
         a_0=(\frac{\pi^2g_{\ast}\epsilon^2\lambda^{\frac{\epsilon-2}{\epsilon}}}
         {90g_sM_P^2\phi_i^{\frac{2\epsilon-4}{\epsilon}}})^{\frac{1}{4}}
         a_{dec}T_{dec}e^{(\frac{\epsilon}{2}-1)N},
    \eea
where we supposed that at the end of inflation, the energy of the
inflaton field transformed to the gauge field completely $i.e$ we
used at the end of the inflation
$3g_sM_P^2H^2=\frac{\pi^2}{30}g_sT_{Reh}$. We also used the fact
that at radiation dominated era $a\sim\frac{1}{T}$.

Here, $g_{\ast}$ is the number of degrees of freedom and $T_{dec}$
is the temperature at decoupling time.

Using $\phi_i\sim b\sqrt{g_s}M_P$ one finds
      \bea    \label{f24}
          h_{\beta}=
          (b^{\frac{-2\epsilon+4}{\epsilon}}
          \frac{\pi^2g_{\ast}\epsilon^2}
         {90g_s})^{-\frac{\epsilon(\beta-1)}
         {2(\epsilon-1)}}(\frac{m}{M_P})^{-2(\beta-1)}
         (\frac{\lambda}{g_s})^{-\frac{\beta-1}{2}\frac{\epsilon-2}{\epsilon-1}}
         (a_{dec}T_{dec})^{-\frac{2(\beta-1)\epsilon}{\epsilon-1}}
         e^{-(\beta-1)\epsilon \frac{\epsilon-2}{\epsilon-1}N}.
         \eea
where $b$ is a constant. One can also find expressions for the
$\eta_i$ and $\eta_f$ using $a_0$ similarly. Inserting
$(\ref{f21})$ and $(\ref{f24})$ in $(\ref{f20})$ one find the
final expression for the $f(\eta)$ in terms of e-folding parameter
$N$.

In the next section we will use the observational data for
evaluate $f(\eta)$ numerically and find the amplitude of the
$B_{dec}$.

\subsection{Estimating the quantities}
First of all, at decoupling time
$\omega_{dec}=k_{dec}/a_{dec}\simeq10^{-33}Gev$, and
 $T_{dec}\sim 10^{-10}Gev$.

Although, it would be nice to study the generation of the magnetic
field during inflation for generic $\epsilon$, but in this paper,
by considering the WMAP analysis \cite{Peiris:2003ff}, we focus on
the smallest $\epsilon$ consistent with a Plank scale inflaton vev
which is $\epsilon\sim 1/20$. For this case we have $b=3/2$. In
this case, the COBE normalization requires a mass hierachy
$\frac{m}{M_P}\simeq 2\times10^{-5}$\cite{Alishahiha:2004eh}.  The
quantity $\lambda/g_s$ corresponds to the number of D$-3$ brane.
For the current case, we have $\frac{\lambda}{g_s}\sim10^{12}$,
which corresponds to an enormous gauge group.

We choose $g_s\sim 1$ because we are at the weakly coupled gravity
regime. We also have $g_{\ast}\simeq106.75$ and
$\frac{1G^2}{8\pi}=1.9089\times10^{-40}Gev^4$.

In this inflationary model, the largest Hubble parameter is
      \bea    \label{f25}
         H\simeq\frac{1}{\sqrt{3g_s}}\frac{m\phi}{M_p}\simeq
         \frac{1}{\sqrt{3}}\frac{m^2}{M_P}.
      \eea
This corresponds to maximum available value for e-foldings
parameter $N\leq 200$. From the usual requirement of solving the
flatness and homogeneity problems, we need at least $60$
e-foldings in total, but the $60$ e-foldings could comes from
multiple passes through the inflationary phase since the scale
invariant spectrum of the CMBR accounts for only about $10$
e-foldings of observations. So $10 \leq N\leq 200$.

We find the amplitude for the minimum available e-foldings $N=10$
using numerical program and obtain
    \bea
          B_{dec}\simeq 3.7\times 10^{-18}G,
    \eea
which is in good agreement with observational data.
\section {Reheating Stage}
The energy density for the scalar field $\phi$ decrease during the
inflation as $\rho \rightarrow \frac{3g_sM_P^2}{\epsilon^2 t^2}$.
In fact, as the theory of inflation and reheating, the energy of
the inflaton field is released to the other field at the end of
the inflation and then the reheating stage started.

At reheating stage the frequency modes greater than the horizon
size is allowed in which one may study the effects of the modes
deep inside the Hubble radius $(k\gg a_eH_e)$. The energy stored
in this modes surpasses the energy of the inflaton field and the
reheating procedure is done.

 In D-cceleration model, the end of inflation is not very clear.
There are different possibility for ending the inflation. In the
other word, when the brane reaches the end of the throat, it can
oscillates of the form discussed in \cite{Kachru:2002kx}, or
annihilates using another anti $D_3$ brane or other novel
phenomena can be happened.

The back reaction of closing of the brane to other branes and
Tachyon production are another effects that one should consider at
the end of such inflationary models.

In the CFT side the reheating stage correspond to the presence of
new light particles $\chi$ which was integrated out. In fact, one
may consider in ${\cal N}=4$ super Yang-Mills theory, $\phi$ as a
Higgs particle breaking the $U(N)$ gauge group to $U(N-1)$.
Reaching the value of $\phi$ to zero means that Higgsed gauge
boson $\chi$ becomes massless and the gauge symmetry restored. So
at reheating stage the effects of virtual particles are important.

So, the end of inflation in this model is rather complicated and
one should consider all the above phenomena for studying reheating
stage.
\section{Conclusion}
Using ADS$/$CFT, an inflationary model was introduced in string
theory in \cite{Silverstein:2003hf} which based on D-cceleration
mechanism. In fact, the author have been noticed the effects of
the virtual gauge bosons at the strongly coupled CFT. In gravity
side, the problem translated to the motion of the probe D$-3$
brane toward the horizon. The causal structure of the theory
causes a speed limit on rolling down of scalar field and power low
evaluation of scale factor at late time. They find for suitably
large inflaton mass $m^2\lambda>g_sM_P^2$ there is such power law
inflation.

The problem of existence of a large scale magnetic field observed
in the galaxies is an unsolved problem for a long time. In fact,
the smallness of the field coherence or the conformal triviality
of the Maxwell equation are some main difficulties to obtain a
good theoretical explanation for this problem. String cosmology
provides a natural framework for studying this problem.

In this paper, we study the production of the primordial magnetic
seed field during inflation considering the above inflationary
model. By coupling the gauge field with scalar field $\phi$ by DBI
action at weakly coupled supergravity, the equation of motion of
gauge field was derived. The generation of mass for the inflaton
field via the coupling of $3-$brane with background $RR$ fluxes,
causes that the conformal invariance of the Maxwell equation
breaks. Then, we write the equation in a standard form and study
the solution of this equation at late time. We finally find the
amplitude for the magnetic field generated during inflation which
is in good agreement with observational data.

There is an important comment. The amount of the $B_k^{dec}$ is
strongly related to the $\epsilon, \lambda/g_s$ and the mass
hierachy $m/M_P$. For example, for the case $\epsilon\simeq 1/20$,
considering the non-Gaussianity with the WMAP analysis, the
starting point for the $\phi$ field is slightly above the Planck
scale $(\phi\sim 1.5\sqrt{g_s}M_P)$. In this case it would entail
a functional fine tuning of the parameters as in early models of
ordinary inflation\cite{Alishahiha:2004eh}. This can be avoided by
studying the generation of magnetic field for the other value of
$\epsilon$ which are in the range $\frac{1}{20}\leq \epsilon\ <
1$.
\section{Appendix}
Consider the following general differential equation
      \bea   \label{h1}
        \ddot{\overrightarrow{A_k}}+F\dot{\overrightarrow{A_k}}
        +\frac{k^2}{a^2\gamma^2}\overrightarrow{A_k}=0,
      \eea
Now, using the change of variables as
$\overrightarrow{A}_k=b^{\frac{-1}{2}}\tilde{\overrightarrow{A_k}}$,
and define conformal time as $\eta=\int a^{-1}dt$ then
      \bea   \label{h2}
          b^{\frac{-1}{2}}a^{-2}\overrightarrow{\tilde{A}}''+
          b^{\frac{-1}{2}}a^{-2}\overrightarrow{A}'[(F-H)a-\frac{b'}{b}]+\hspace{6cm}\cr
          \overrightarrow{\tilde{A}}[-\frac{F}{2}a^{-1}b^{\frac{-3}{2}}b'+
          b^{\frac{-3}{2}}a^{-2}(\frac{H}{2}ab'-\frac{b''}{2})+
          \frac{3}{4}a^{-2}b^{\frac{-5}{2}}b'^2-\frac{k^2}{a^2(\gamma^2)}]=0.\hspace{0.5cm}
      \eea
The second term can be cancelled provided that
      \bea \label {h3}
         (F-H)a=\frac{b'}{b}=a\frac{\dot{b}}{b}.
      \eea
and finally one obtains
     \bea    \label{h4}
         \tilde{\overrightarrow{A}}''+\tilde{\overrightarrow{A}}([-\frac{a^2}{4}(F^2-H^2)-
         \frac{a^2}{2}(\dot{F}-\dot{H})]+k^2(\gamma^2)^{-1})=0.
     \eea
So, for the case
$\gamma(\phi)=\frac{1}{\sqrt{1-f(\phi)\dot{\phi}^2}}$ one finds
     \bea  \label{h5}
         \tilde{\overrightarrow{A}}''+[k^2-f(\eta)]\tilde{\overrightarrow{A}}=0,
     \eea
where
     \bea   \label{h6}
          f(\eta)=k^2f\dot{\phi}^2+\frac{a^2}{4}[(F^2-H^2)+2(\dot{F}-\dot{H})].
     \eea
This equation was called "Generalized Emden-Fowler" equation.
\section*{Acknowledgment}
I would like to thank Dr M. Alishahiha to bring me the problem and
valuable discussion. I would like to thank M.R. Garousi, A.
Ghodsi, A. Mosaafa, S. Rahvar, S. Sheikh-Jabbari , E. Silvrstein,
A. Tavanfar, M. Torabian, M. Torkiha and D. Tong for comments. I
would especially like to thank S. Movahed for valuable helps in
numerical analysis of the problem.


\begin{thebibliography}{99}
\bibitem{Bassett:2000aw}
  B.~A.~Bassett, G.~Pollifrone, S.~Tsujikawa and F.~Viniegra,
  ``Preheating as cosmic magnetic dynamo,''
  Phys.\ Rev.\ D {\bf 63} (2001) 103515
  [arXiv:astro-ph/0010628].

\bibitem{Davis:1999bt}
  A.~C.~Davis, M.~Lilley and O.~Tornkvist,
  ``Relaxing the Bounds on Primordial Magnetic Seed Fields,''
  Phys.\ Rev.\ D {\bf 60} (1999) 021301
  [arXiv:astro-ph/9904022].

\bibitem{Maroto:2001ki}
  A.~L.~Maroto,
  ``Cosmological magnetic fields induced by metric perturbations after
  inflation,''
  arXiv:hep-ph/0111268.

\bibitem{Turner:1987bw}
  M.~S.~Turner and L.~M.~Widrow,
  ``Inflation Produced, Large Scale Magnetic Fields,''
  Phys.\ Rev.\ D {\bf 37} (1988) 2743.

\bibitem{Betschart:2005iu}
  G.~Betschart, C.~Zunckel, P.~Dunsby and M.~Marklund,
  ``Primordial magnetic seed field amplification by gravitational waves,''
  arXiv:gr-qc/0503006.

\bibitem{Dolgov:1993vg}
  A.~Dolgov,
  ``Breaking of conformal invariance and electromagnetic field generation in
  the universe,''
  Phys.\ Rev.\ D {\bf 48} (1993) 2499
  [arXiv:hep-ph/9301280].

\bibitem{Battaner:2000kf}
  E.~Battaner and H.~Lesch,
  ``On the physics of primordial magnetic fields,''
  arXiv:astro-ph/0003370.

\bibitem{Mazumdar:2000jc}
  A.~Mazumdar and M.~M.~Sheikh-Jabbari,
  ``Noncommutativity in space and primordial magnetic field,''
  Phys.\ Rev.\ Lett.\  {\bf 87} (2001) 011301
  [arXiv:hep-ph/0012363].

\bibitem{Gasperini:1995dh}
  M.~Gasperini, M.~Giovannini and G.~Veneziano,
  ``Primordial magnetic fields from string cosmology,''
  Phys.\ Rev.\ Lett.\  {\bf 75} (1995) 3796
  [arXiv:hep-th/9504083].

\bibitem{Garousi:2004hy}
  M.~R.~Garousi, M.~Sami and S.~Tsujikawa,
  ``Generation of electromagnetic fields in string cosmology with a massive
  scalar field on the anti D-brane,''
  Phys.\ Lett.\ B {\bf 606} (2005) 1
  [arXiv:hep-th/0405012].

\bibitem{Bamba:2004cu}
  K.~Bamba and J.~Yokoyama,
  ``Large-scale magnetic fields from dilaton inflation in noncommutative
  spacetime,''
  Phys.\ Rev.\ D {\bf 70} (2004) 083508
  [arXiv:hep-ph/0409237].

\bibitem{Davis:2005ih}
  A.~C.~Davis and K.~Dimopoulos,
  ``Cosmic superstrings and primordial magnetogenesis,''
  arXiv:hep-ph/0505242.

\bibitem{Giddings:2001yu}
  S.~B.~Giddings, S.~Kachru and J.~Polchinski,
  ``Hierarchies from fluxes in string compactifications,''
  Phys.\ Rev.\ D {\bf 66} (2002) 106006
  [arXiv:hep-th/0105097].

\bibitem{Silverstein:2003hf}
  E.~Silverstein and D.~Tong,
  ``Scalar speed limits and cosmology: Acceleration from D-cceleration,''
  Phys.\ Rev.\ D {\bf 70}, 103505 (2004)
  [arXiv:hep-th/0310221].

\bibitem{Alishahiha:2004eh}
  M.~Alishahiha, E.~Silverstein and D.~Tong,
  ``DBI in the sky,''
  Phys.\ Rev.\ D {\bf 70}, 123505 (2004)
  [arXiv:hep-th/0404084].

\bibitem{Aharony:1999ti}
  O.~Aharony, S.~S.~Gubser, J.~M.~Maldacena, H.~Ooguri and Y.~Oz,
  ``Large N field theories, string theory and gravity,''
  Phys.\ Rept.\  {\bf 323}, 183 (2000)
  [arXiv:hep-th/9905111].

\bibitem{Kachru:2003aw}
  S.~Kachru, R.~Kallosh, A.~Linde and S.~P.~Trivedi,
  ``De Sitter vacua in string theory,''
  Phys.\ Rev.\ D {\bf 68} (2003) 046005
  [arXiv:hep-th/0301240].

\bibitem{Kachru:2003sx}
  S.~Kachru, R.~Kallosh, A.~Linde, J.~Maldacena, L.~McAllister and S.~P.~Trivedi,
  ``Towards inflation in string theory,''
  JCAP {\bf 0310} (2003) 013
  [arXiv:hep-th/0308055].
  G.~Dvali and S.~Kachru,
  arXiv:hep-th/0309095.
  J.~P.~Hsu, R.~Kallosh and S.~Prokushkin,
  JCAP {\bf 0312} (2003) 009
  [arXiv:hep-th/0311077].
  A.~Buchel and R.~Roiban,
  Phys.\ Lett.\ B {\bf 590} (2004) 284
  [arXiv:hep-th/0311154].
  H.~Firouzjahi and S.~H.~H.~Tye,
  Phys.\ Lett.\ B {\bf 584} (2004) 147
  [arXiv:hep-th/0312020].
  L.~Pilo, A.~Riotto and A.~Zaffaroni,
  JHEP {\bf 0407} (2004) 052
  [arXiv:hep-th/0401004].
  A.~Linde,
  Phys.\ Scripta {\bf T117} (2005) 40
  [arXiv:hep-th/0402051].
  M.~R.~Garousi, M.~Sami and S.~Tsujikawa,
  Phys.\ Rev.\ D {\bf 70} (2004) 043536
  [arXiv:hep-th/0402075].
  R.~H.~Brandenberger,
  arXiv:astro-ph/0411671.
  C.~P.~Burgess, J.~M.~Cline, H.~Stoica and F.~Quevedo,
  JHEP {\bf 0409} (2004) 033
  [arXiv:hep-th/0403119].
  A.~Buchel and A.~Ghodsi,
  Phys.\ Rev.\ D {\bf 70} (2004) 126008
  [arXiv:hep-th/0404151].
  H.~Yavartanoo,
  arXiv:hep-th/0407079.
  A.~Ghodsi and A.~E.~Mosaffa,
  Nucl.\ Phys.\ B {\bf 714} (2005) 30
  [arXiv:hep-th/0408015].
  X.~g.~Chen,
  Phys.\ Rev.\ D {\bf 71} (2005) 063506
  [arXiv:hep-th/0408084].
  R.~H.~Brandenberger,
  arXiv:astro-ph/0411671.
  X.~g.~Chen,
  arXiv:hep-th/0501184.

\bibitem{Maldacena:1997re}
  J.~M.~Maldacena,
  ``The large N limit of superconformal field theories and supergravity,''
  Adv.\ Theor.\ Math.\ Phys.\  {\bf 2} (1998) 231
  [Int.\ J.\ Theor.\ Phys.\  {\bf 38} (1999) 1113]
  [arXiv:hep-th/9711200].

\bibitem{Spergel:2003cb}
  D.~N.~Spergel {\it et al.}  [WMAP Collaboration],
  ``First Year Wilkinson Microwave Anisotropy Probe (WMAP) Observations:
  Determination of Cosmological Parameters,''
  Astrophys.\ J.\ Suppl.\  {\bf 148} (2003) 175
  [arXiv:astro-ph/0302209].

\bibitem{Peiris:2003ff}
  H.~V.~Peiris {\it et al.},
  ``First year Wilkinson Microwave Anisotropy Probe (WMAP) observations:
  Implications for inflation,''
  Astrophys.\ J.\ Suppl.\  {\bf 148} (2003) 213
  [arXiv:astro-ph/0302225].

\bibitem{Kachru:2002kx}
  S.~Kachru and L.~McAllister,
  ``Bouncing brane cosmologies from warped string compactifications,''
  JHEP {\bf 0303} (2003) 018
  [arXiv:hep-th/0205209].
\end{thebibliography}
\end{document}